\documentclass[showkeys,eqsecnum,showpacs,preprintnumbers,amssymb,aps]{revtex4}
\usepackage{graphicx}
\usepackage{slashbox}
\usepackage{dcolumn}
\usepackage{bm}
\usepackage{latexsym,epsfig}

\begin{document}

\preprint{\today}

\title{{\it Ab initio} studies of electron correlation effects in the atomic parity violating amplitudes in Cs and Fr}

\author{B. K. Sahoo \footnote{B.K.Sahoo@rug.nl}}
\affiliation{KVI, University of Groningen, NL-9747 AA Groningen, The Netherlands}

\begin{abstract}
We have studied the correlation effects in Cs and Fr arising from the interplay of the residual
Coulomb interaction to
all orders and the neutral weak interaction which gives rise to the parity violating electric
dipole transition to first order, within the framework of the relativistic coupled-cluster theory which 
circumvents the constrain of explicitly summing over the intermediate states. We observe that, the  
contributions arising from the perturbed doubly excited states are quite significant and hence, any 
calculation should not be considered accurate unless it includes the perturbed double excitations 
comprehensively. In this article, we have reported a comparative study of various results related to the 
parity violation in Cs and Fr.
\end{abstract}


\maketitle

\section{Introduction}
One of the most challenging contemporary problems in physics is the search for possible new physics beyond the 
well known standard model (SM) of elementary particles \cite{mohapatra,barr,liu,ginges}. Apart from using 
gigantic accelerators at high-energy scales, it is also possible to use high precision, albeit, low-energy 
table top atomic experiments, such as, the measurement of atomic parity violation, in combination with 
accurate relativistic many-body calculations of the atomic parity nonconserving transition amplitude, 
$E_{1_{PNC}}$ to achieve this goal
\cite{ginges,bouchiat}. Some of the prominent signatures of physics beyond 
the SM which can be inferred from these atomic experiments are: a tight limit on the mass of extra 
Z-bosons, precise value of the Weinberg angle, limit to the radiative 
corrections for the electron-nuclear weak interactions etc \cite{ginges}.
As some of these SM results are known to high precision, they demand similar sub-one percent accuracies in 
both the measurements and the atomic calculations.
As the interaction Hamiltonian for the atomic parity violation (APV) due to the nuclear spin-independent (NSI) 
electron-nucleus interactions is proportional to $Z^3$ where $Z$ being the atomic number \cite{bouchiat}, 
heavy atomic systems are chosen for the study of APV effects.
A series of APV experiments on a number of atomic systems, including those of Cs \cite{wood} and Tl 
\cite{vetter} have been carried out. However, the high accuracy of $\sim 0.35\%$ has been achieved only for Cs.
Furthermore, a number of ab initio calculations of APV amplitudes in Cs have been carried out
using a variety of many-body approaches. Some results based
on the relativistic coupled-cluster (RCC) theory are also available for Cs,
however, their accuracies are somewhat uncertain, since most of them have used the 
sum-over-states approach which considers contributions from the core orbitals 
approximately and accounts for only a selected number of excited states whose 
contributions are dominant. In addition, the doubly excited intermediate atomic states and the normalization
of the RCC wave functions are treated only approximately. 
We have developed a technique in the frame work of RCC theory that circumvents 
these drawbacks and it has been employed earlier in the calculation of the 
$E1_{PNC}$ amplitudes in Ba$^+$ \cite{bijayaba} and Ra$^+$ \cite{bijayara} in 
which we have demonstrated that the accuracies of $< 1\%$ and $< 3\%$, 
respectively, were possible.

In this work, we employ the new RCC approach, mentioned above, to study various
correlation effects in the parity violating amplitudes in Cs and Fr. We report 
a comparative study of their results along with those reported previously.

\section{Theory of APV}

\begin{figure}[h]
\includegraphics[width=5.0cm]{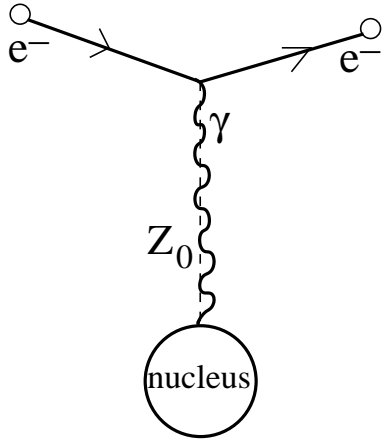}
\caption{Diagrammatic representation of the electron-nucleus interactions due to the electromagnetic (with the exchange of photon $\gamma$ denoted by solid curved line) and the weak (with the exchange of heavy boson $Z_0$ denoted by straight dashed line) interactions.}
\label{fig1}
\end{figure}
The dominant interaction in an atom is the electromagnetic interaction which, as well known, conserves parity. 
However, there is a non-zero probability of the interaction between the electrons and the nucleus of an atom 
due to the weak force with the exchange of a $Z_0$ boson, as shown in 
Fig. \ref{fig1}, which violates parity. The interaction Hamiltonian between the electrons and the
nucleus due to the weak interaction can have two components: one, 
vector--axial-vector and the other, axial-vector--vector currents. The latter depends on
the nuclear spin and most of its contribution cancels out except from the odd
nucleon and hence, it is relatively smaller in magnitude \cite{bouchiat} than
the former; the NSI component. In this work, we shall consider the APV effect due to the NSI
component alone.

The APV interaction Hamiltonian due to the NSI component is given by,
\begin{eqnarray}
H_{APV}^{NSI} &=& \frac {G_F}{2\sqrt{2}} Q_w(N) \gamma_5^e \rho_N^e(r),
\label{eqn1}
\end{eqnarray}
\noindent
where $G_F$ is the Fermi constant, $\rho_N^e(r)$ is the electron density over
the nucleus and $\gamma_5^e(=i\gamma_0^e\gamma_1^e\gamma_2^e\gamma_3^e)$, 
is the product of the four Dirac matrices that involve electron spin, and 
$Q_w(N)$ is the nuclear weak charge, which is equal to $2(Z c_{p} + N c_{d})$ 
where $c_p$ and $c_n$ denote the electron-proton and the electron-neutron 
coupling constants for the atomic ($Z$) and neutron ($N$) numbers,
respectively. The values of these coupling constants predicted by the SM, in 
the lowest order of electroweak interaction (at the tree level), are given by
\begin{eqnarray}
c_{p} = \frac{1}{2}(1-4sin^2\theta_W) \approx 0.04, \hspace*{1.0cm} c_n =-\frac{1}{2} ,
\label{eqn2}
\end{eqnarray}
where $\theta_W$ is the Weinberg angle and its measured value is $\sim sin^2\theta_W \approx 0.23$ \cite{anthony}. Substituting these values in $Q_w(N)$,
we get $Q_w(N)=-N+Z(1-sin^2\theta_W)$ which is proportional to $N$.
Hence, the perturbation due to $H_{APV}^{NSI}$ is generally expressed in the 
scale of $Q_w(N)/N$.
Since $H_{APV}^{NSI}$ does not commute with the parity operator, its inclusion
with the atomic Hamiltonian of the electromagnetic interaction, which 
commutes with the parity operator, mixes the opposite parity states of same angular momentum.
The strength of this interaction is sufficiently weak, which justifies its consideration as a 
first-order perturbation.

The Dirac-Coulomb (DC) Hamiltonian used, here, in the calculation of the atomic wave functions
of definite parity is given by
\begin{eqnarray}
H^{DC} = \sum_i \left [ c \alpha \cdot p_i + (\beta -1)c^2 + V_{nuc}(r_i) \right ] + \sum_{i > j} \frac{1}{r_{ij}} , \nonumber \\
\label{eqn3}
\end{eqnarray}
where $c$ is the velocity of light, $\alpha$ and $\beta$ are the Dirac matrices
(note that $\gamma^e=\beta \alpha$) and $V_{nuc}(r)$ is the nuclear potential.

The atomic wave functions ($|\Psi_n^{(0)} \rangle$) corresponding to 
$H^{DC}$ can be considered as the unperturbed wave functions. The total wave function of a system including the first order correction due to the interaction Hamiltonian $H_{APV}^{NSI}$ is given by
\begin{equation}
|\Psi_n \rangle = |\Psi_n^{(0)} \rangle + G_F |\Psi_n^{(1)} \rangle ,
\label{eqn4}
\end{equation}
where $|\Psi_n^{(1)} \rangle$ is the first order perturbed wave function of its unperturbed valence 
state $|\Psi_n^{(0)} \rangle$ and $G_F$ is used as a coupling constant.

Electric dipole (E1) transitions between the states of same parity are forbidden due to the electromagnetic 
selection rules. However, an E1 transition between the states of mixed parity, mixed due to $H_{APV}^{NSI}$ 
interaction, is possible and the corresponding transition amplitude can be expressed as
\begin{eqnarray}
E1_{PNC} &=& \frac {\langle \Psi_f| D |\Psi_i \rangle } {\sqrt{\langle \Psi_f|\Psi_f \rangle \langle \Psi_i|\Psi_i \rangle }},
\label{eqn5}
\end{eqnarray}
where $D=e\,r$ is the E1 operator, the subscripts $i$ and $f$ denote initial
and final valence orbitals, respectively.

Expanding the total wave function as given in Eq. (\ref{eqn4}) and 
retaining the terms only up to first order in $G_F$, we get
\begin{eqnarray}
E1_{PNC} &=& G_F \frac {\langle \Psi_f^{(0)}| D |\Psi_i^{(1)} \rangle + \langle \Psi_f^{(1)}| D |\Psi_i^{(0)} \rangle } {\sqrt{\langle \Psi_f^{(0)}|\Psi_f^{(0)} \rangle \langle \Psi_i^{(0)}|\Psi_i^{(0)} \rangle }} \nonumber \\
&=& G_F \sum_{I \ne i} \frac {\langle \Psi_f^{(0)}| D |\Psi_I^{(0)} \rangle \langle \Psi_I^{(0)}| H_{APV}^{NSI} |\Psi_i^{(0)} \rangle } {E_i - E_I} \nonumber \\
 && + G_F \sum_{J \ne f} \frac {\langle \Psi_f^{(0)}| H_{APV}^{NSI}|\Psi_J^{(0)} \rangle \langle \Psi_J^{(0)}| D |\Psi_i^{(0)} \rangle } {E_f - E_J} ,
\label{eqn6}
\end{eqnarray}
where the subscripts $I$ and $J$ represent the intermediate unperturbed states. We have used, here, the 
explicit form for the first order wave function given by
\begin{eqnarray}
|\Psi_n^{(1)} \rangle &=& \frac{1}{G_F} \sum_{I \ne n} |\Psi_I^{(0)} \rangle \frac
{ \langle \Psi_I^{(0)}| H_{APV}^{NSI} |\Psi_n^{(0)} \rangle } {E_n - E_I}.
\label{eqn7}
\end{eqnarray}

 An important question we address in this paper is: How significant 
are the contributions from those states which were considered approximately in 
the sum-over-states approach and how they vary with the size of the systems?
We would address this by carrying out a comparative study of 
$E1_{PNC}$ results in two systems, namely Cs and Fr, of different atomic sizes.
In order for the contributions of these
higher excited states to be included, it is necessary to solve the first order perturbation equation
directly. In other words, it is necessary to solve the equation 
following equation
\begin{eqnarray}
(H^{DC} - E^{(0)})|\Psi_n^{(1)}\rangle =& \frac{1}{G_F} (E^{(1)} - H_{APV}^{NSI}) |\Psi_n^{(0)}\rangle ,
\label{eqn8}
\end{eqnarray}
where $E^{(1)} (=\langle \Psi_n^{(0)}| H_{APV}^{NSI} |\Psi_n^{(0)} \rangle)$
is the first order correction to $E^{(0)}$ which, however, vanishes in the 
present case.

\section{Application of RCC theory to APV}
The RCC method, which is equivalent to all order perturbation theory, has
been used in the recent past and accurate results have been reported for many 
single valence systems \cite{bijayaba,bijayara,csur,sahoo}. In the RCC 
framework, the wave function of a single valence atom can be expressed as
\begin{eqnarray}
| \Psi_n^{(0)} \rangle &=& = e^{T^{(0)}} \{ 1+S_n^{(0)} \} | \Phi_n \rangle,
\label{eqn9}
\end{eqnarray}
where $| \Phi_n \rangle$ is the reference state constructed from the
Dirac-Fock (DF) wave function $| \Phi_0 \rangle$ of the closed-shell
configuration by appending the valence electron n, that is, 
$|\Phi_n \rangle= a_n^{\dagger} | \Phi_0 \rangle$ where $a_n^{\dagger}$
represents a creation operator which creates the valence electron $n$. Here $T^{(0)}$ and
$S_n^{(0)}$
are the RCC excitation operators which excite electrons from $| \Phi_0 \rangle$
and $|\Phi_n \rangle$, respectively, due to the residual Coulomb interactions. The corresponding excitation amplitudes
are obtained by solving the following equations
\begin{eqnarray}
\langle \Phi^L |\{\widehat{H_N^{DC} e^{T^{(0)}}}\}|\Phi_0 \rangle &=& 0 \label{eqn10} \\
\langle \Phi_n^L|\{\widehat{H_N^{DC} e^{T^{(0)}}}\}S_n|\Phi_n\rangle &=& - \langle \Phi_n^L|\{\widehat{H_N^{DC} e^{T^{(0)}}}\}|\Phi_n\rangle + \langle \Phi_n^L|S_n|\Phi_n\rangle \Delta E_n^{(0)} ,
\label{eqn11}
\end{eqnarray}
with the superscript $L(=1,2)$ representing the singly and doubly excited
states from the corresponding reference states and the wide-hat symbol over
$H_N^{DC} e^{T^{(0)}}$ represent the linked terms of normal order atomic
Hamiltonian $H_N^{DC}$
and RCC operator $T^{(0)}$. In the CCSD (CC with single and double excitations) approximation, the corresponding RCC operators are defined by
\begin{eqnarray}
T^{(0)} &=& T_1^{(0)} + T_2^{(0)}
\label{eqn12}
\end{eqnarray}
and
\begin{eqnarray}
S_n^{(0)} &=& S_{1n}^{(0)} + S_{2n}^{(0)}
\label{eqn13}
\end{eqnarray}
for the closed-shell and single valence open-shell systems, respectively. The quantity 
$\Delta E_n^{(0)}$ in the above expression is the electron affinity energy (or negative of the ionization potential (IP)) for the valence electron which is evaluated by
\begin{eqnarray}
 \Delta E_n^{(0)} = \langle \Phi_n|\{\widehat{H_N^{DC} e^{T^{(0)}}}\} \{1+S_n^{(0)}\} |\Phi_n\rangle.
\label{eqn14}
\end{eqnarray}
In addition to having considered full singles and doubles in the CCSD equations given in Eq. (\ref{eqn11}), we have also included the contributions
from the important triple excitations perturbatively (known in the literature as CCSD(T) method) by defining
\begin{eqnarray}
S_{3n}^{(0),pert} &=&  \widehat{H_N^{DC} T_2^{(0)}} + \widehat{H_N^{DC} S_{2n}^{(0)}},
\label{eqn15}
\end{eqnarray}
where the superscript $pert$ denotes the perturbation and their contributions to $\Delta E_n^{(0)}$ are  evaluating as
\begin{eqnarray}
\Delta E_n^{(0),trip} &=& \widehat{T_2^{(0)\dagger} S_{3n}^{(0),pert}} .
\label{eqn16}
\end{eqnarray}
After solving for the amplitudes of $T^{(0)}$, we solve Eqs. (\ref{eqn11}) and (\ref{eqn14}) simultaneously and obtain the amplitudes of $S_n^{(0)}$ operators.

Now, the total atomic wave function in the presence of $H_{APV}^{NSI}$ is expressed, in the RCC ansatz,  as
\begin{eqnarray}
| \Psi_n \rangle &=& e^{T^{(0)}+ G_F T^{(1)}} \{ 1+S_n^{(0)} + G_F S_n^{(1)} \} | \Phi_n \rangle,
\label{eqn17}
\end{eqnarray}
where $T^{(1)}$ and $S_n^{(1)}$ are the first order perturbed amplitudes corresponding to the unperturbed RCC 
operators $T^{(0)}$ and $S_n^{(0)}$, respectively. On expanding the above equation keeping the terms  only up 
to first order in $G_F$ yields
\begin{eqnarray}
| \Psi_n \rangle &=& e^{T^{(0)}} \{ 1+ S_n^{(0)}+ G_F T^{(1)} ( 1+S_n^{(0)}) + G_F S_n^{(1)} \} | \Phi_n \rangle . 
\label{eqn18}
\end{eqnarray}

Comparing the above equation with Eq. (\ref{eqn4}), we get
\begin{eqnarray}
| \Psi_n^{(1)} \rangle &=& e^{T^{(0)}} \{ T^{(1)} ( 1+S_n^{(0)}) + S_n^{(1)} \} | \Phi_n \rangle .
\label{eqn19}
\end{eqnarray}

In order to calculate $| \Psi_n^{(1)} \rangle$ as a solution of Eq. (\ref{eqn8}) in the RCC theory, we solve the excitation operator amplitudes of $T^{(1)}$
and $S_n^{(1)}$ using the following equations
\begin{eqnarray}
\langle \Phi^L |\{\widehat{H_N^{DC} e^{T^{(0)}}} T^{(1)} \}|\Phi_0 \rangle &=&  - \langle \Phi^L | \widehat{H_{APV}^{NSI} e^{T^{(0)}}} |\Phi_0 \rangle  
\label{eqn20}
\end{eqnarray}
and
\begin{eqnarray}
\langle \Phi_n^L|\{\widehat{H_N^{DC} e^{T^{(0)}}}\} S_n^{(1)} |\Phi_n\rangle &=& - \langle
\Phi_n^L|\{\widehat{H_N^{DC} e^{T^{(0)}}} T^{(1)} ( 1+S_n^{(0)}) \nonumber \\
&& + \widehat{H_{APV}^{NSI} e^{T^{(0)}}} ( 1+S_n^{(0)})
 \}|\Phi_n\rangle 
+ \langle \Phi_n^L|S_n^{(1)} |\Phi_n\rangle \Delta E_n , \ \ \
\label{eqn21}
\end{eqnarray}
after solving Eq. (\ref{eqn10}) and Eq. (\ref{eqn11}), respectively. In the
above expression, notation $\widehat{H_{APV}^{NSI} e^{T^{(0)}}}$ is used for
the connecting terms between $H_{APV}^{NSI}$ and $T^{(0)}$. To keep the level of approximation uniform through 
out, both $T^{(1)}$ and
$S_n^{(1)}$ are truncated at single and double excitations by defining
\begin{eqnarray}
T^{(1)} &=& T_1^{(1)} + T_2^{(1)}
\label{eqn22}
\end{eqnarray}
and
\begin{eqnarray}
S_n^{(1)} &=& S_{1n}^{(1)} + S_{2n}^{(1)}
\label{eqn23}
\end{eqnarray}
where $T_1^{(1)}$ and $S_{1n}^{(1)}$ correspond to the perturbed single excitations and $T_2^{(1)}$ and $S_{2n}^{(1)}$ correspond to the perturbed double excitations, from closed- and open-shells, respectively. Since both the perturbed single and double excitation amplitudes are solved simultaneously, certain correlation effects due to the perturbed double excitations also reflect indirectly in the contributions of the perturbed single excitations.

After obtaining the unperturbed and the perturbed RCC operator amplitudes in both the closed-shell and one-valence open-shell atoms, we proceed to calculate the $E1_{PNC}$ amplitude as
\begin{eqnarray}
E1_{PNC} &=& G_F \frac {\langle \Psi_n^{(0)} | D | \Psi_n^{(1)} \rangle
+ \langle \Psi_n^{(1)} | D | \Psi_n^{(0) } \rangle} { \langle \Psi_n^{(0) }|\Psi_n^{(0) } \rangle} \nonumber \\
&=& G_F \frac{ \langle \Phi_n |\{1+S_n^{(0)\dagger}\} \overline{D} \{T_1^{(1)} (1+S_n^{(0)} ) + S_{1n}^{(1)} \} | \Phi_n \rangle } {\langle \Phi_n | \{1+S_n^{(0) \dagger}\} \overline{N}_0 \{1 +S_n^{(0)} \} | \Phi_n \rangle } \nonumber \\ &&
+ G_F \frac { \langle \Phi_n |\{S_n^{(1)\dagger} + (1+S_n^{(0) \dagger}) T^{(1)\dagger} \} \overline{D} \{1 +S_n^{(0)} \} | \Phi_n \rangle } { \langle \Phi_n | \{1+S_n^{(0) \dagger}\} \overline{N}_0 \{1 +S_n^{(0)} \} | \Phi_n \rangle },
\end{eqnarray}
\label{eqn24}
where we define $\overline{D}=(e^{T^{(0) \dagger}} D e^{T^{(0)}})$ and
$\overline{N}_0 = e^{T^{(0)\dagger}} e^{T^{(0)}}$. The non-truncative series for
$\overline{D}$ and $\overline{N}_0$ are expanded using the Wick's
generalized theorem and are truncated when the terms are below fifth order of
the Coulomb interaction. The core-valence and valence correlation
contributions are obtained from $DT_1^{(1)}$ and $D (T_1^{(1)}\{1+S_n^{(0)} \}+S_{1n}^{(1)})$, respectively, along with their conjugate terms.

Corrections due to the normalization of the wave functions are accounted by
evaluating
\begin{eqnarray}
Norm &=& \left [ \langle \Psi_n^{(0)} | D | \Psi_n^{(1)} \rangle + \langle \Psi_n^{(1)} | D | \Psi_n^{(0)} \rangle \right ] \left \{ \frac {N_n}{1+N_n} \right \}, \nonumber \\
\label{eqn25}
\end{eqnarray}
where $N_n=\langle \Phi_n | \{1+S_n^{(0)\dagger}\} \overline{N}_0 \{1 +S_n^{(0)} \}  | \Phi_n \rangle $.

 Although, the CCSD(T) method described here, accounts for the contributions from the important
unperturbed triple excitations it fails to include the direct triple excitation contributions to the $E1_{PNC}$ calculations. To account for, at least, the
lowest order direct triple excitation contributions (minimum up to fourth order in Coulomb interaction),
we construct them with the open-shell RCC operators perturbatively as follows
\begin{eqnarray}
S_{nab}^{pqr,(0)} &=&  \frac{\widehat{H_N^{DC} T_2^{(0)}} + \widehat{H_N^{DC} S_{2n}^{(0)}}}{\epsilon_n + \epsilon_a + \epsilon_b - \epsilon_p - \epsilon_q - \epsilon_r},
\end{eqnarray}
and
\begin{eqnarray}
S_{nab}^{pqr,(1)} &=&  \frac{\widehat{H_N^{DC} T_2^{(1)}} + \widehat{H_N^{DC} S_{2n}^{(1)}}}{\epsilon_n + \epsilon_a + \epsilon_b - \epsilon_p - \epsilon_q - \epsilon_r},
\end{eqnarray}
where $\epsilon_i$ is the single particle energy of an orbital $i$. These 
operators are finally considered as parts of $S_n^{(0)}$ and $S_n^{(1)}$ in 
Eq. (\ref{eqn24}). This can be called as lo-CCSDvT approximation, which also 
accounts important lower order valence triple excitation effects in the final
property calculations.

\section{Results and Discussions}
\subsection{Orbitals generation}
We have used Gaussian type functions
\begin{eqnarray}
F^{GTO}(r_i) = r^{n_{\kappa}} e^{-\alpha_i r_i^2}
\label{eqn26}
\end{eqnarray}
to construct the DF orbitals where $\alpha_i$ is an arbitrary
parameter which has to be chosen and $r_i$ represents a radial grid given by
\begin{eqnarray}
r_i = r_0 \left [ e^{h(i-1)} - 1 \right ]
\label{eqn27}
\end{eqnarray}
where the step size $h$ is taken to be $0.03$, the radial grid is increased up to
$i=750$, $r_0$ is $ 2 \times 10^{-6}$
in atomic units and $n_{\kappa}$ is the radial quantum number of the orbitals.
Here $\alpha_i$s are chosen to satisfy the even tempering condition
\begin{eqnarray}
\alpha_i = \alpha_0 \beta^{i-1}
\label{eqn28}
\end{eqnarray}
and we have chosen different $\alpha_0$ and $\beta$ values for different
symmetries ($l$) (known as even tempered basis) and they are given in
Table \ref{tab1}.
\begin{table}[h]
\caption{Used $\alpha_0$ and $\beta$ parameters for different symmetries ($l$) to construct DF orbitals using GTOs in Cs and Fr.}
\begin{tabular}{l|ccccc}
\hline
\hline
\backslashbox{$\alpha_0/\beta$}{$l$} & 0  & 1 & 2 & 3 & 4 \\
\hline
 & & & \\
$\alpha_0$ & 0.00190 & 0.001825 & 0.00183 & 0.00185 & 0.00187 \\
$\beta$ & 2.91 & 2.94 & 2.94 & 3.05 & 3.09 \\
\hline
\hline
\end{tabular}
\label{tab1}
\end{table}

The finite size of the nucleus in these
systems is accounted by assuming a two-parameter Fermi-nuclear-charge
distribution for evaluating the electron density over the nucleus as given by
\begin{equation}
\rho(r_i) = \frac {\rho_0} {1 + e^{(r_i-c)/a}},
\label{eqn29}
\end{equation}
where $\rho_0$ is the density for the point nuclei, $c$ and $a$ are the
half-charge radius and skin thickness of the
nucleus. These parameters are chosen as
\begin{equation}
a = 2.3/4(ln3)
\label{eqn30}
\end{equation}
and
\begin{equation}
c = \sqrt{ \frac{5}{3} r_{rms}^2 - \frac{7}{3} a^2 \pi^2},
\label{eqn31}
\end{equation}
where $r_{rms}$ is the root mean square radius of the corresponding nuclei
which is determined by
\begin{equation}
r_{rms} = 0.836 A^{1/3} + 0.570
\label{eqn32}
\end{equation}
in $fm$ for the atomic mass $A$.

\begin{table}[t]
\caption{Comparison of $E1_{PNC}$ results of $^{133}$Cs and $^{223}$Fr due to electron-electron Coulomb interactions from various methods in $\times 10^{-11} iea_0 (-Q_W/N)$.}
\begin{center}
\begin{tabular}{cccc}
\hline
\hline
 $^{133}$Cs & $^{223}$Fr & Method & Reference \\
$6s \ ^2S_{1/2} \rightarrow 7s \ ^2S_{1/2}$ & $7s \ ^2S_{1/2} \rightarrow 8s \ ^2S_{1/2}$ & & \\
\hline
 & & \\
$-0.7236$  & $-13.4825$ & DF & This work \\
$-0.8941$  & $-14.5134$ & CCSD & This work \\
$-0.8889$  & $-14.4219$  & CCSD(T) & This work \\
$-0.8892$  & $-14.4106$ & lo-CCSDvT & This work \\
$-0.904(9)$  & & LCCSD(T) $+$ SL $+$ SS & \cite{blundell} \\
$-0.9078$  & & BO $+$ GFCP & \cite{dzuba} \\
$-0.904$  & $-15.72$ & CI & \cite{shabaev} \\
$-0.902(4)$  & & CCSD(T) & \cite{das} \\
$-0.9001$  & $-15.229$ & BO $+$ GFCP & \cite{dzuba1} \\
$-0.8998(25)$  &  & CCSDvT $+$ SS & \cite{porsev} \\
 & $-15.9$ & BO $+$ GFCP & \cite{dzuba2} \\
 & $-15.41(17)$ & LCCSD $+$ RPA $+$ ExpEng $+$ SS & \cite{safronova}\\
 & & \\
\hline
\hline
\end{tabular}
\end{center}
\begin{tabular}{l}
NOTE: Contributions from Breit interaction, QED correction and nuclear effects are not considered here.
\end{tabular}
\begin{tabular}{llll}
Abbreviations & SS & : & sum-over-states \\
 & SL & : & scaling \\
 & BO & : & Brueckner orbitals \\
 & GFCP & : & Green function technique for all order correlation potential \\
 & CI & : & configuration interaction method \\
 & ExpEng & : & experimental energy \\
 & RPA & : & random phase approximation \\
\end{tabular}
\label{tab2}
\end{table}

\begin{table}[t]
\caption{Contributions to the $E1_{PNC}$ calculations in Cs and Fr using CCSD(T) method (in $\times 10^{-11} iea_0 (-Q_W/N)$).}
\begin{tabular}{lclc}
\hline
\hline
(A) \ Cs & & \\
Initial pert. terms & & Final pert. terms &  \\
($6s ^2S_{1/2}^{(1)} \rightarrow 7s ^2S_{1/2}^{(0)}$) & & ($6s ^2S_{1/2}^{(0)} \rightarrow 7s ^2S_{1/2}^{(1)}$) & \\
\hline
DF result \\
$D H_{APV}^{NSI} $ & $-1.0460$ & $ H_{APV}^{NSI} D $ & 0.3224 \\
\hline
 & & \\
$ T_1^{(1)\dagger}D $ & $0.0374$ &  $D T^{(1)} $ & $-0.0355$ \\
$\overline{D}_1 S_{1i}^{(1)}$ & $-1.7467$ & $S_{1f}^{(1)\dagger} \overline{D}_1 $ & $0.1841$ \\
$\overline{D}_1 S_{2i}^{(1)}$ & $-0.0046$ & $S_{2f}^{(1)\dagger} \overline{D}_1 $ & $0.0226$  \\
$S_{1f}^{(0)\dagger} \overline{D}_1 S_{1i}^{(1)}$ & $0.4061$ & $S_{1f}^{(1)\dagger} \overline{D}_1  S_{1i}^{(0)} $ & $0.2141$ \\
$S_{2f}^{(0)\dagger} \overline{D}_1 S_{1i}^{(1)}$ & $-0.0055$ & $S_{1f}^{(1)\dagger} \overline{D}_1  S_{2i}^{(0)} $ & $0.0301$ \\
$S_{1f}^{(0)\dagger} \overline{D}_1 S_{2i}^{(1)}$ & $-0.0026$ & $S_{2f}^{(1)\dagger} \overline{D}_1  S_{1i}^{(0)} $ & $-0.0028$ \\
$S_{2f}^{(0)\dagger} \overline{D}_1 S_{2i}^{(1)}$ & $0.0011$ & $S_{2f}^{(1)\dagger} \overline{D}_1  S_{2i}^{(0)} $ & $-0.0073$ \\
$Others$ & $-0.0045$ &  & 0.0054 \\
$Norm$ & 0.0301 & & $-0.0105$ \\
 & & \\
\hline
\hline
(B) \ Fr & & \\
Initial pert. terms & & Final pert. terms & \\
($7s ^2S_{1/2}^{(1)} \rightarrow 8s ^2S_{1/2}^{(0)}$) & & ($7s ^2S_{1/2}^{(0)} \rightarrow 8s ^2S_{1/2}^{(1)}$) & \\
\hline
 DF result \\
$D H_{APV}^{NSI} $ & $-19.4661$ &  $ H_{APV}^{NSI} D $ & 5.9836  \\
\hline
  & & \\
$ T_1^{(1)\dagger}D $ & $0.7147$ &  $D T^{(1)} $ & $-0.6876$ \\
$\overline{D}_1 S_{1i}^{(1)}$ & $-31.7107$ & $S_{1f}^{(1)\dagger} \overline{D}_1 $ & $4.0538$ \\
$\overline{D}_1 S_{2i}^{(1)}$ & $-0.0314$ & $S_{2f}^{(1)\dagger} \overline{D}_1 $ & $0.5930$ \\
$S_{1f}^{(0)\dagger} \overline{D}_1 S_{1i}^{(1)}$ & $7.8001$ & $S_{1f}^{(1)\dagger} \overline{D}_1  S_{1i}^{(0)} $ & $4.1072$ \\
$S_{2f}^{(0)\dagger} \overline{D}_1 S_{1i}^{(1)}$ & $-0.0423$ & $S_{1f}^{(1)\dagger} \overline{D}_1  S_{2i}^{(0)} $ & $0.6580$ \\
$S_{1f}^{(0)\dagger} \overline{D}_1 S_{2i}^{(1)}$ & $-0.0615$ & $S_{2f}^{(1)\dagger} \overline{D}_1  S_{1i}^{(0)} $ & $-0.0911$ \\
$S_{2f}^{(0)\dagger} \overline{D}_1 S_{2i}^{(1)}$ & $0.0033$ & $S_{2f}^{(1)\dagger} \overline{D}_1  S_{2i}^{(0)} $ & $-0.1501$ \\
$Others$ & $-0.0152$ & & 0.0769 \\
$Norm$ & 0.6043 & & $-0.2433$ \\
\hline
\hline
\end{tabular}
\label{tab3}
\end{table}

\subsection{Results}
In Table \ref{tab2}, we give  the $E1_{PNC}$ amplitude results obtained from
various calculations. From this work, we present the results using  the DF, CCSD,
CCSD(T) and lo-CCSDvT methods for both Cs and Fr. In the same table, we also
compare our results with previously reported results using various
many-body methods. Our Cs result matches reasonably well with the other
calculations, but our Fr result differs significantly. We give 
below individual contributions from various RCC terms and express them in terms
of correlation diagrams and level of excitations in order to facilitate the
readers to understand the role of various correlation effects and intermediate
states. Briefly, we discuss here the methods used in the other calculations.
The difference between our DF and lo-CCSDvT results gives an idea about the
amount of total correlation effects through the present method in these 
calculations. Considering the same DF wave functions, we have also employed
CCSD and CCSD(T) methods where we find that the CCSD results are larger in magnitude
than the CCSD(T) results. However, the $E1_{PNC}$ result increases in magnitude for Cs 
in the lo-CCSDvT approximation, but it decreases for Fr; indicating that
the triple excitation effects are stronger in Fr. From a comparison between
these results, we observe that the dominant triple excitations effects arise
through the CCSD(T) method.

\subsection{Discussions}
Here we discuss briefly different reported calculations of the above results at 
the DC approximation. About two decades ago, Blundell et al \cite{blundell}
had employed the linearized
CCSD(T) method (LCCSD(T) method) to evaluate the unperturbed wave functions
in Cs and then they had used a sum-over-states approach to evaluate the $E1_{PNC}$
amplitude for the $6s \ ^2S_{1/2} \rightarrow 7s \ ^2S_{1/2}$ transition in Cs.
However, they had scaled their wave functions to fit the calculated energies of
different states with the experimental results which reproduced many atomic
properties quite accurately, but that does not show that the method they had used is capable
of producing accurate ab initio results.
Shabaev et al \cite{shabaev} have also obtained results using a CI method with 
a local form of the DF wave functions. Their Cs result matches with the result 
of Blundell et al. In our previous work on Cs \cite{das}, we had also 
calculated this quantity by considering the same $\alpha_0$ and $\beta$ parameters
for orbitals of all symmetries (known as universal basis). With the new 
parameters, the Gaussian basis orbitals produce better wave functions in the nuclear region which are 
verified by studying the hyperfine interactions that will be reported elsewhere.
The convergence of the RCC amplitudes are better than $10^{-8}$ in the 
present case than $10^{-6}$ in \cite{das} and it gives a slightly different 
result. Dzuba et el have carried out a few calculations of these quantities 
using Brueckner orbitals using a Green function technique (Feynman diagram
approach) that takes into account various classes of correlation effects to all orders and avoids 
the sum-over-states approach \cite{dzuba,dzuba1,dzuba2}. Their results also
differ from each other and in some cases with others as can be seen in Table \ref{tab1}.
The most recent calculation on Cs is reported by Porsev et al
\cite{porsev} using the RCC method that includes all single and double excitations
with all valence triple excitations (CCSDvT method). However, they have finally
used a sum-over-states approach to calculate the $E1_{PNC}$ amplitude of the
$6s \ ^2S_{1/2} \rightarrow 7s \ ^2S_{1/2}$ transition. They give contributions
from $6p_{1/2}$ to $9p_{1/2}$ singly excited states as "Main", which contributes
$-0.8823(18)$, and the remaining contributions as "Tail", which is
obtained using other many-body methods as $-0.0175(18)$, (results are always 
given in $\times 10^{-11} iea_0 (-Q_W/N)$ here onwards) at the DC approximation.
 In contrast to their approach, we have used the lo-CCSDvT method, but have
included contributions from the core and doubly excited states in a manner similar 
to that of the singly excited states.

\begin{table}[t]
\caption{Important contributions to the $E1_{PNC}$ calculations in Cs from
various intermediate states ($I$ and $J$ notations are used as per Eq. 
(\ref{eqn6}) in $\times 10^{-11} iea_0 (-Q_W/N)$).}
\begin{tabular}{lcclcc}
\hline
\hline
$6s ^2S_{1/2}^{(1)} \rightarrow 7s ^2S_{1/2}^{(0)}$ & & & $6s ^2S_{1/2}^{(0)} \rightarrow 7s ^2S_{1/2}^{(1)}$ & & \\
  & $I$ & Results & & $J$ & Results \\
\hline
 & & \\
$T_1^{(1)\dagger}D$ & & & $D T^{(1)}$ & & \\
& $4p_{1/2} \rightarrow 7s$ & $0.0005$ & & $4p_{1/2} \rightarrow 6s$ & $-0.0005$ \\
& $5p_{1/2} \rightarrow 7s$ & $0.0368$ & & $5p_{1/2} \rightarrow 6s$ & $-0.0349$ \\
$\overline{D}_1 S_{1i}^{(1)}$ & & & $S_{1f}^{(1)\dagger} \overline{D}_1$ & & \\
& $6s \rightarrow 6p_{1/2}$ & $-1.9195$ & & $7s \rightarrow 6p_{1/2}$ & $1.7648$ \\
& $6s \rightarrow 7p_{1/2}$ & $0.1532$ & & $7s \rightarrow 7p_{1/2}$ & $-1.5022$ \\
& $6s \rightarrow 8p_{1/2}$ & $0.0347$ & & $7s \rightarrow 8p_{1/2}$ & $-0.0902$ \\
& $6s \rightarrow 9p_{1/2}$ & $-0.0084$ & & $7s \rightarrow 10p_{1/2}$ & $0.0093$ \\
& $6s \rightarrow 10p_{1/2}$ & $-0.0051$ & & & \\
$S_{1f}^{(0)\dagger} \overline{D}_1 S_{1i}^{(1)}$ & & & $S_{1f}^{(1)\dagger} \overline{D}_1 S_{1i}^{(0)}$ & & \\
& $6s \rightarrow 6p_{1/2}$ & $0.1716$ & & $7s \rightarrow 6p_{1/2}$ & $0.2556$ \\
& $6s \rightarrow 7p_{1/2}$ & $0.1302$ & & $7s \rightarrow 7p_{1/2}$ & $0.1213$ \\
& $6s \rightarrow 8p_{1/2}$ & $0.0258$ & & $7s \rightarrow 8p_{1/2}$ & $-0.0611$ \\
& $6s \rightarrow 9p_{1/2}$ & $0.0735$ & & $7s \rightarrow 9p_{1/2}$ & $-0.0946$ \\
& $6s \rightarrow 10p_{1/2}$ & $0.0053$ & & $7s \rightarrow 10p_{1/2}$ & $-0.0055$\\
$\overline{D}_1 S_{2i}^{(1)}$ & & & $S_{2f}^{(1)\dagger} \overline{D}_1$ & & \\
& $6s5p_{1/2} \rightarrow 7s6s$ & $-0.0019$ & & $7s5p_{3/2} \rightarrow 6s9s$ & $0.0002$ \\
& $6s5p_{1/2} \rightarrow 7s8s$ & $-0.0001$ & & $7s5p_{3/2} \rightarrow 6s10s$ & $0.0009$ \\
& $6s5p_{3/2} \rightarrow 7s8s$ & $0.0001$ & & $7s5p_{1/2} \rightarrow 6s6d_{3/2}$ & $0.0001$ \\
& $6s5p_{1/2} \rightarrow 7s9s$ & $-0.0009$ & & $7s5p_{3/2} \rightarrow 6s6d_{3/2}$ & $-0.0003$ \\
& $6s5p_{3/2} \rightarrow 7s9s$ & $0.0009$ & & $7s5p_{3/2} \rightarrow 6s7d_{3/2}$ & $-0.0001$ \\
& $6s5p_{1/2} \rightarrow 7s10s$ & $-0.0014$ & & $7s5p_{1/2} \rightarrow 6s8d_{3/2}$ & $0.0006$ \\
& $6s5p_{3/2} \rightarrow 7s10s$ & $0.0001$ & & $7s5p_{3/2} \rightarrow 6s8d_{3/2}$ & $-0.0003$ \\
& $6s5s \rightarrow 7s9p_{1/2}$ & $0.0001$ & & $7s5p_{1/2} \rightarrow 6s9d_{3/2}$ & $0.0019$ \\
& $6s5s \rightarrow 7s10p_{1/2}$ & $0.0002$ & & $7s5p_{3/2} \rightarrow 6s9d_{3/2}$ & $0.0003$ \\
& $6s5p_{1/2} \rightarrow 7s6d_{3/2}$ & $0.0006$ & & $7s5p_{1/2} \rightarrow 6s10d_{3/2}$ & $0.0005$ \\
& $6s5p_{3/2} \rightarrow 7s6d_{3/2}$ & $0.0005$ & & $7s5p_{3/2} \rightarrow 6s10d_{3/2}$ & $0.0001$ \\
& $6s5p_{1/2} \rightarrow 7s7d_{3/2}$ & $0.0003$ & & $7s5p_{3/2} \rightarrow 6s5d_{5/2}$ & $0.0044$ \\
& $6s5p_{3/2} \rightarrow 7s7d_{3/2}$ & $0.0002$ & & $7s5p_{3/2} \rightarrow 6s6d_{5/2}$ & $0.0018$ \\
& $6s5p_{1/2} \rightarrow 7s8d_{3/2}$ & $0.0013$ & & $7s5p_{3/2} \rightarrow 6s7d_{5/2}$ & $0.0008$ \\
& $6s5p_{3/2} \rightarrow 7s8d_{3/2}$ & $0.0011$ & & $7s5p_{3/2} \rightarrow 6s8d_{5/2}$ & $0.0041$ \\
& $6s5p_{1/2} \rightarrow 7s9d_{3/2}$ & $0.0006$ & & $7s5p_{3/2} \rightarrow 6s9d_{5/2}$ & $0.0048$ \\
& $6s5p_{3/2} \rightarrow 7s9d_{3/2}$ & $0.0005$ & & $7s5p_{3/2} \rightarrow 6s10d_{5/2}$ & $0.0003$ \\
& $6s5p_{1/2} \rightarrow 7s10d_{3/2}$ & $-0.0004$ & & $7s4d_{3/2} \rightarrow 6s8f_{5/2}$ & $0.0003$ \\
& $6s5p_{3/2} \rightarrow 7s10d_{3/2}$ & $-0.0002$ & & $7s4d_{3/2} \rightarrow 6s9f_{5/2}$ & $0.0005$ \\
& $6s5p_{3/2} \rightarrow 7s8d_{5/2}$ & $-0.0003$ & & $7s4d_{5/2} \rightarrow 6s8f_{7/2}$ & $0.0004$ \\
& $6s5p_{3/2} \rightarrow 7s9d_{5/2}$ & $-0.0013$ & & $7s4d_{5/2} \rightarrow 6s8f_{7/2}$ & $0.0004$ \\
& $6s5p_{3/2} \rightarrow 7s10d_{5/2}$ & $-0.0012$ & & $7s4d_{5/2} \rightarrow 6s9f_{7/2}$ & $0.0009$\\
& $6s4d_{3/2} \rightarrow 7s8f_{5/2}$ & $-0.0001$ \\
& $6s4d_{3/2} \rightarrow 7s9f_{5/2}$ & $-0.0004$ \\
& $6s4d_{5/2} \rightarrow 7s8f_{7/2}$ & $-0.0005$ \\
\hline
\hline
\end{tabular}
\label{tab4}
\end{table}

\subsection{Cs vs. Fr}
\begin{figure}[t]
\includegraphics[width=8.5cm,clip=true]{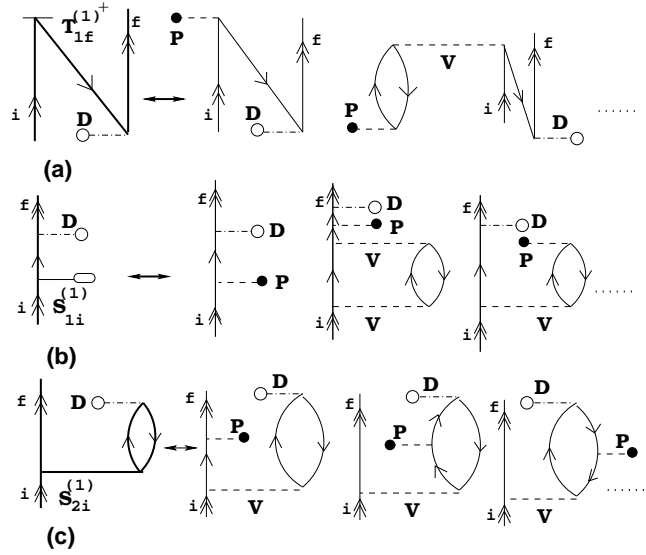}
\caption{Break-down of the perturbed RCC diagrams into the lower-order 
many-body theory (MBPT) diagrams. Here
$i$ and $f$ represent the initial and final orbitals, $P$ is the PNC operator
and $V$ is the two-body part of the Coulomb interaction Hamiltonian. Lines
going up and down with single arrows represent virtual and hole (core) orbitals
and lines with double arrows represent attached valence orbitals to the
closed-shell DF wave function.}
\label{fig2}
\end{figure}

In Table \ref{tab3}, we present the individual contributions from different
RCC terms in our CCSD(T) method to the $E1_{PNC}$ calculations for the $6s ^2S_{1/2}\rightarrow 7s ^2S_{1/2}$ and $7s ^2S_{1/2}\rightarrow 8s ^2S_{1/2}$
transitions in Cs and Fr, respectively. It is evident that the trends of the contributions from the different RCC terms are similar for both Cs and Fr.
The important contributions come from three terms:
$\overline{D}_1 S_{1i}^{(1)}$, $\overline{D}_1 S_{2i}^{(1)}$ and
$D T_1^{(1)}$ and their corresponding conjugate terms where $\overline{D}_1$ is the effective one-body terms of $\overline{D}$ and bare operator $D$ is its
lowest order term. Other terms correspond to higher order RCC terms, but
they are not small. Contributions given as $Others$ represent
the RCC terms that come from the effective two-body terms of $\overline{D}$
after contracting with the open-shell RCC terms. We give diagrammatic
representations of the above three RCC terms in Fig \ref{fig2} (without
their conjugate terms) and their 
lowest order terms.  From this figure, it can be noticed that
$D T_1^{(1)}$ and $\overline{D}_1 S_{1i}^{(1)}$ contain the lowest order DF
contributions from the core (hole) and virtual orbitals, respectively; hence
$\overline{D}_1 S_{1i}^{(1)}$ always gives the largest contribution. Again,
the perturbed states arising through the ground state contribute predominantly
while contributions from the perturbed excited states are comparatively smaller
than those corresponding to the ground state but with the opposite signs. The final results are the
outcome of these cancellations. The most core correlation effects are coming
from $D T_1^{(1)}$ and its conjugate terms while there are small contributions that
come from $\overline{D} T_2^{(1)}$ and its conjugate terms which are included 
in "Others". As seen these correlation effects mostly cancel out with their
corresponding conjugate terms in both the systems. The contributions from singly excited states 
can be estimated by summing the total contributions from 
$\overline{D}_1 S_{1i}^{(1)}$, $S_{1f}^{(0)\dagger}\overline{D}_1 S_{1i}^{(1)}$,
 $S_{2f}^{(0)\dagger} \overline{D}_1 S_{1i}^{(1)}$ and their conjugate RCC
terms. The remaining contributions are come from the doubly excited states
apart from the normalization corrections to the RCC wave functions. After
accounting the corresponding corrections from the normalization of the 
wave functions to various RCC operators, we get 
contributions as $0.0018$, $-0.8980$ and $0.0073$ from the core correlation,
singly excited states and doubly excited states, respectively, in Cs.
Similarly, we get the contributions as $0.0271$, $-14.7551$ and $0.3158$ from 
the core correlation, singly excited states and doubly excited states,
respectively, in Fr. Clearly, the core correlation and doubly excited states
contributions in Fr are large. In Cs and Fr, the doubly excited states
contributions are around 1\% and 2\%, respectively, with opposite signs. Hence,
these contributions should also be accounted accurately for high
precision results.

 To understand more clearly about the role of different excited states
that play important roles in the $E1_{PNC}$ result of Cs through our RCC
approach which have also been studied by other methods, we consider some
of the important RCC terms and analyze their contributions through various
excitations level. These results are given explicitly in Table \ref{tab4}.
As noticed, only $4p_{1/2}$ and $5p_{1/2}$ core orbitals are important to
account the core correlation effects; this is because the single particle 
energy levels between these orbitals with the valence electrons are small.
The singly excited states from $6p_{1/2}$ to $9p_{1/2}$ contribute
predominantly, but there are also significant contributions from some of the continuum states. 
The continuum orbital, $10p_{1/2}$, whose density in the nuclear region is large.
There are various doubly excited states that play important roles, mainly from 
the $6s$k$p$ to $7s$l$d$ and $7s$k$p$ to $6s$l$d$ excitations for any arbitrary 
principal quantum numbers k and l, in calculating the above quantities but
mostly they cancel each other. Therefore, a suitable choice of basis 
functions and an accurate many-body method are necessary for accounting 
these contributions precisely.

\begin{figure}[t]
\includegraphics[width=8.0cm,clip=true]{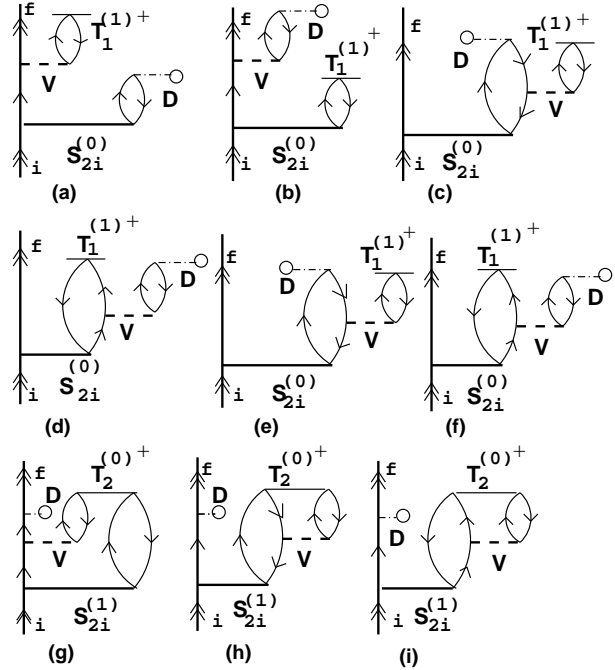}
\caption{Some of the important contributing diagrams from the triple excitations which contribute significantly to the lo-CCSDvT method along with their conjugate terms.}
\label{fig3}
\end{figure}
 To compare the role of the correlation effects in calculating $E1_{PNC}$
amplitudes in Cs and Fr, we compare the DF results and then their final
results. It is about 20\% in Cs, but it is only about 7\% in Fr. The small
correlation effects in Fr is because of strong cancellations between the 
contributions from the initial and final perturbed states. Again,
it is interesting to note that the CCSD results were larger than the CCSD(T)
results suggesting that there were cancellations from the triple excitations.
Using the lo-CCSDvT method, these results increase in Cs slightly but 
cancel out in Fr.  The triple excitation effects in Fr are strong
enough to be taken seriously into account for high precision results.
In Fig. \ref{fig3}, we give some of the important RCC diagrams (without
their conjugate terms) that give large contributions through the lo-CCSDvT 
method. In the future, we would like to study other relevant properties in order to
determine the accuracy of our calculated $E1_{PNC}$ amplitudes in these systems.

\section{Conclusion}
We have applied the relativistic coupled-cluster method to calculate the atomic
parity violating effects in Cs and Fr which involve the interplay of the long 
range electrostatic interactions and the short range weak interactions. The 
present approach considers the Coulomb interaction up to all orders and the 
weak interaction between the nucleus and the electrons up to first order. From 
the detailed analysis of different contributions, it is clear that the doubly 
excited states play an important role in obtaining precise results. These
contributions are quite large in magnitude, however, with opposite signs,
than those of the singly excited states in Fr, resulting in a large
cancellation in the final results. They are about 1\% in Cs,  for which the
experimental accuracies have been claimed to be around 0.35\%. This suggests
that, it requires a method like our present approach to consider them 
accurately for high precision calculations.

\section{Acknowledgment}
I am grateful to B. P. Das for useful discussions and H. S. Nataraj for
his critical reading of the manuscript. 
This work was supported by NWO under the VENI program with Project No.
680-47-128. I thank the C-DAC TeraFlop Super Computing facility, Bangalore,
India for the cooperation to carry out these calculations on its computers.

\end{document}